\documentclass[review, sort&compress]{elsarticle}

\usepackage{hyperref}
\usepackage{pgf,tikz}
\usepackage{mathrsfs}
\usetikzlibrary{arrows}
\usepackage{graphicx}
\graphicspath{ {figures/} }
\usepackage{caption}
\usepackage{subcaption}
\usepackage{amsmath}
\allowdisplaybreaks

\journal{SI: Computer and Mathematics with Applications}

\begin{document}

\begin{frontmatter}

\title{Implementation of the compact interpolation within the octree based Lattice Boltzmann
solver Musubi}


\author[siegen,aachen]{Jiaxing Qi\corref{cor1}}
\ead{qijiaxing@gmail.com}
\author[siegen]{Harald Klimach}
\ead{harald.klimach@uni-siegen.de}
\author[siegen]{Sabine Roller}
\ead{sabine.roller@uni-siegen.de}
\cortext[cor1]{Corresponding author}

\address[siegen]{University of Siegen, H\"olderlinstr. 3, Siegen 57068, Germany}
\address[aachen]{RWTH Aachen University, Templergraben 55, Aachen 52062, Germany}

\begin{abstract}
A sparse octree based parallel implementation of the lattice Boltzmann method
for non-uniform meshes is presented in this paper.
To couple grids of different resolutions, a second order accurate compact
interpolation is employed and further extended into three dimensions.
This compact interpolation requires only four source elements from the adjacent
level for both two- and three dimensions.
Thus, it reduces the computational and communication overhead in parallel executions.
Moreover, the implementation of a weight based domain decomposition algorithm
and level-wise elements arrangement are explained in details.
The second order convergence of both velocity and strain rate are validated
numerically in the Taylor-Green vortex test case.
Additionally, the laminar flow around a cylinder at $Re=20,100$ and
around a sphere at $Re=100$ is investigated.
Good agreement between simulated results and those from literature is observed,
which provides further evidence for the accuracy of our method.
\end{abstract}


\end{frontmatter}

\newcommand{\vect}[1]{\boldsymbol{#1}}
\newcommand{\density}{\rho}
\newcommand{\vel}{u}
\newcommand{\vecvel}{\mathbf{\vel}}
\newcommand{\vecx}{\mathbf{x}}
\newcommand{\vecc}{\mathbf{c}}
\newcommand{\pdf}{\emph{pdf}}
\newcommand{\pressure}{p}
\newcommand{\sxx}{S_{xx}}
\newcommand{\syy}{S_{yy}}
\newcommand{\szz}{S_{zz}}
\newcommand{\sxy}{S_{xy}}
\newcommand{\syz}{S_{yz}}
\newcommand{\sxz}{S_{xz}}
\newcommand{\der}[2]{\frac{\partial #1}{\partial{#2}}}
\newcommand{\dt}{\Delta t}
\newcommand{\dx}{\Delta x}
\newcommand{\musubi}{\emph{Musubi}}
\newcommand{\treeid}{\emph{treeID}}
\newcommand{\gfc}{\emph{GhostFromCoarser}}
\newcommand{\gff}{\emph{GhostFromFiner}}
\newcommand{\etal}{\emph{et al.}}

\section{Introduction}

The Lattice Boltzmann Method (LBM) is a powerful numerical scheme for low Mach
number flow simulations.
Instead of discretizing the incompressible Navier-Stokes equations directly,
it is based on mesoscopic kinetic models and is capable of incorperating
other physical models to simulate complex flows, such as multiphase and
multispecies flows~\cite{chen1998lattice,bernsdorf2006concurrent,aidun2010lattice,zudrop2014lattice}.
The LBM algorithm is simple and requires only direct neighbor information, making it
a good candidate for large scale parallel simulations~\cite{zeiser2009benchmark,schonherr2011multi,eitel2013lattice,hasert2014complex}.
But its reliance on uniform grids limits its numerical efficiency when there are
multiple spatial scales within a single simulation.
For example the direct computation of sound generating flows along with the
propagation of generated waves into the far field.
To tackle such problems, LBM with local mesh refinement techniques have been
developed in recent years.
This broadened the applicability of LBM to many more practical areas of
interest.

Filippova \etal{}~\cite{filippova1998grid} proposed an interpolation based
local mesh
refinement method where the nonquilibrium part of the particle distribution
function (\pdf) between fine and coarse mesh is scaled to keep the viscosity
continuous.
The equilibrium part on the other hand is directly interpolated between mesh
resolutions.
This idea has been adopted and further investigated in several studies since
then~\cite{lin2000lattice,yu2002multi,dupuis2003theory,tolke2006adaptive,
  yu2006multi,chen2011lattice,eitel2011lattice,lagrava2012advances,
  hasert2012:aeroacoust,neumann2013:a-dynamic-,eitel2013lattice,
  fakhari2014finite,fakhari2015numerics}.

In contrast to pointwise interpolation schemes, a grid refinement based on
a volumetric formulation has been proposed by Chen and Chen \etal{}~\cite{chen1998volumetric,chen2006grid}.
In their method, the exact conservation of mass and momentum between regions with
different element size is achieved.
In another strategy, macroscopic quantities or moments are used as
interpolation quantities, as shown in~\cite{geier2009bubble,tolke2009second,hasert2013thesis}.
In the current work, the mesh refinement technique based on the second order
accurate compact interpolation as proposed by Geier \etal{}~\cite{geier2009bubble}
is extended into three dimensions and its convergence and accuracy is validated.
Only four source elements are needed to perform the quadratic interpolation
in both two and three dimensions.
Thus, the numerical efficiency of the multilevel simulation is increased and
data communication is reduced to a minimum, which is especially important
for parallel runs.

The mesh refinement technique has been implemented in the parallel
LB flow solver \musubi{}, which is part of the APES simulation framework.
It is built around a central octree library~\cite{Harald:2012aa},
where the common tree operations are encapsulated in the MPI-parallel environments.
Grids can be refined locally in arbitrarily complex geometries without any
restrictions on parallel executions.
Its applications in various scientific and engineering fields and
computational performance on several High Performance Computing (HPC) systems have been
reported in~\cite{hasert2011towards,zimny2013:a-multisca,hasert2014complex,qi2015aeroacoustic}.

The remainder of this article is structured as follows.
Section~\ref{sec:method} provides a brief introduction to the LBM algorithm together
with the compact interpolation scheme for mesh refinement.
Section~\ref{sec:implementation} explains some details on the octree based
implementation, of which the domain decomposition algorithm and level-wise
elements arrangement is highlighted.
The convergence and accuracy of our approach is presented
in Section~\ref{sec:results} through the test cases of the Taylor-Green vortex,
the flow around a cylinder (2D) and the flow around a sphere (3D).
Finally, Section~\ref{sec:conclusion} gives a conclusion and an outlook to
future works.

\section{Numerical Method}
\label{sec:method}

In this section, we first briefly review the basic LBM algorithm. Then we
explain the compact interpolation for the mesh refinement in details.

\subsection{Lattice Boltzmann Method}

Instead of discretising the incompressible Navier-Stokes equations directly,
the LBM describes fluid flows at the mesoscopic level.
The state of fluid is represented by the particle distribution function
$f_i(\vecx,t)$,
meaning the probability of finding a particle with the velocity
$\vecc{}_{i}$ at a spatial position $\vecx$ and time $t$.
The velocity space is discretised into a finite number of $q$ directions,
resulting in the so-called $DdQq$ lattice, where $d$ is the number of spatial
dimensions.
Common models include the $D2Q9$, $D3Q15$, $D3Q19$ and $D3Q27$ formulations.
In this study, the $D2Q9$ and $D3Q27$ stencils are used, the latter has
been shown to provide better consistency and axisymmetric solutions for
axisymmetric problems than 3D stencils with less directions~\cite{augier2014rotational,silva2014truncation}.

The evolution of $f_{i}$ in time is given by the LB equation
\begin{equation}
  f_{i}(\vecx+\vecc{}_i\Delta t,t+\Delta t) =
    f_{i}(\vecx,t) - \Omega_{i}(f_{i}(\vecx,t)),
  \label{eq:lbe}
\end{equation}
where $\Omega_{i}(f_{i}(\vecx,t))$ is the collision operator that describes
the change in $f_{i}$ resulting from the collision of particles.
Several collision operators have been proposed to recover the correct
hydrodynamic behavior on the macroscopic scale.
Among them, the BGK model~\cite{qian1992bgk}
enjoys its popularity by using a single relaxation time for all the moments.
It is given by
\begin{equation}
  f_{i}(\vecx+\vecc{}_i\Delta t,t+\Delta t) =
    f_{i}(\vecx,t) - \omega(f_{i}(\vecx,t)-f^{eq}_{i}(\vecx,t)),
\label{eq:bgk}
\end{equation}
where $\omega$ controls the relaxation frequency.
The equilibirium distribution function $f_i^{eq}$ is obtained by truncating the
Maxwell-Boltzmann equilibrium distribution up to the second order velocity term,
\begin{equation}
  f_i^{eq} = t_i \rho \left( 1 + \frac{\vecc\cdot\vecvel}{c_s^2}
                               + \frac{(\vecc\cdot\vecvel)^2}{2 c_s^4}
                               + \frac{\vecvel\cdot\vecvel}{2 c_s^2}
                      \right).
\label{eq:feq}
\end{equation}
where $c_s$ is the speed of sound.

The macroscopic quantities, density $\rho$ and momentum $\rho\vecvel$, are
defined as particle velocity moments of the distribution function, $f_i$,
\begin{equation}
  \rho = \sum_{i=1}^{q} f_{i},
  \qquad
  \rho\vecvel = \sum_{i=1}^{q} \vecc_{i}f_{i}
\end{equation}
and the pressure $p$ is obtained from $p=c_s^2\rho$,
where $c_s = ({\Delta x}/{\Delta t})/{\sqrt{3}}$ is the speed of sound.
The kinematic viscosity $\nu$ can be obtained by the relaxation parameter
$\omega$ as
\begin{equation} 
  \nu = \frac{1}{3} \left( \frac{1}{\omega} - \frac{1}{2} \right)
        \frac{\Delta x^2}{\Delta t}
\label{eq:nu}
\end{equation}
The strain rate tensor $S_{\alpha\beta}$ is locally available and can be
computed from the non-equilibirium part of distribution function by
\begin{equation}
S_{\alpha\beta} = - \frac{3\omega}{2\rho}\sum_i c_{i\alpha}c_{i\beta}f_i^{neq}
\end{equation}

\subsection{Local Mesh Refinement}

The locally refined mesh used in this work is built upon the sparse octree
library TreElM~\cite{Harald:2012aa}.
Within an octree, the size of a mesh element is related to its refinement level $L$ by
\begin{equation}
  \Delta x = H / 2^L
\end{equation}
where $H$ is the maximum length of the enclosing computational domain.
A volumetric based view is adopted where distribution functions are assumed
at the barycenters of the elements of the octree.
The TreElM implementation can deal with arbitrary refinement jumps at level
interfaces.
However, this is achieved by recursively applying single level interpolations.
Without loss of generality, we therefore, can consider just the interpolation
between mesh resolutions with $L$ and $L+1$.
Usually, it is also good to maintain this relation for the numerical scheme,
and the mesh is generated under the constraint to contain only interfaces with
at most one level difference.
The relation $m$ of the coarse element edge length $\Delta x_c$ to the one of
the fine elements $\Delta x_f$ thereby is fixed to $m = \Delta x_c / \Delta x_f = 2$.
To achieve consistency in the Mach number across levels, the so-called acoustic
scaling is applied (i.e. $\Delta t \propto \Delta x$).
Thus, when one iteration is performed on the coraser level, two iterations
have to be performed on the finer level.
Besides the Mach number, viscosity and Reynolds number are kept constant across
levels by adjusting the relaxation parameter
\begin{equation}
  \frac{1}{\omega_f} - \frac{1}{2} =
  m \left( \frac{1}{\omega_c} - \frac{1}{2} \right)
  \label{eq:fneq_scale}
\end{equation}

When neighbor elements are on a different level, a direct advection can not be
performed.
Instead, ghost elements that act as a placeholder to provide interpolated values
for normal fluid elements are introduced at the level interfaces.
There are two types of ghost elements within current implementation:
\gfc{} which are filled with data from a
coarser level and \gff{} which are filled with data from a finer level.

Previous studies~\cite{he1996some,dupuis2003theory,rheinlander2005consistent,chen2011lattice,
   fakhari2014finite,fakhari2015numerics}
have indicated that a quadratic or cubic interpolation is necessary
to maintain the second order convergence of the LBM algorithm.
However, Eitel \etal{}~\cite{eitel2011lattice,eitel2013lattice} have obtained
good results by using linear interpolations in combination with a subgrid-scale
model.
In the current study, the velocity is interpolated quadraticly by using a
compact stencil and the velocity gradient information (i.e. strain rate tensor)
that is locally available, while the pressure and $f^{neq}$ is interpolated
linearly.
The idea of compact interpolation, first proposed by
Geier~\cite{geier2009bubble}, is to use a minimum of only four source elements,
providing a high degree of locality.
The algorithm for two dimensions can be found in~\cite{geier2009bubble}.
In present study, we extended this algorithm to three dimensions and explain
it in the following.

Assuming a second order spatial polynomial for velocity:
\begin{equation}
\resizebox{.9\hsize}{!}{$
  \vecvel{(x,y,z)} =
   \left( \begin{array}{lll}
      a_{0} + a_{x}x    + a_{y}y    + a_{z}z
            + a_{xx}x^2 + a_{yy}y^2 + a_{zz}z^2
            + a_{xy}xy  + a_{yz}yz  + a_{xz}xz \\
      b_{0} + b_{x}x    + b_{y}y    + b_{z}z
            + b_{xx}x^2 + b_{yy}y^2 + b_{zz}z^2
            + b_{xy}xy  + b_{yz}yz  + b_{xz}xz \\
      c_{0} + c_{x}x    + c_{y}y    + c_{z}z
            + c_{xx}x^2 + c_{yy}y^2 + c_{zz}z^2
            + c_{xy}xy  + c_{yz}yz  + c_{xz}xz
   \end{array} \right)
$}
\end{equation}
There are 30 unknown coefficients in total, thus 30 equations are required.
Firstly, we choose four source elements, which have the local coordinates:
$H(0,0,0)$, $K(1,1,0)$, $M(1,0,1)$, $N(0,1,1)$.
Each source element provides three velocity components, thus 12 equations:
\begin{subequations} \label{eq:compact_3D_u}
\begin{align}
  \vecvel (0,0,0) &=
   \left( \begin{array}{rrr}
      a_{0}\\
      b_{0}\\
      c_{0}
   \end{array} \right) \\
  \vecvel (1,1,0) &=
   \left( \begin{array}{rrr}
      a_{0} + a_x + a_y + a_{xx} + a_{yy} + a_{xy}\\
      b_{0} + b_x + b_y + b_{xx} + b_{yy} + b_{xy}\\
      c_{0} + c_x + c_y + c_{xx} + c_{yy} + c_{xy}
   \end{array} \right) \\
  \vecvel (1,0,1) &=
   \left( \begin{array}{rrr}
      a_{0} + a_x + a_z + a_{xx} + a_{zz} + a_{xz}\\
      b_{0} + b_x + b_z + b_{xx} + b_{zz} + b_{xz}\\
      c_{0} + c_x + c_z + c_{xx} + c_{zz} + c_{xz}
   \end{array} \right) \\
  \vecvel (0,1,1) &=
   \left( \begin{array}{rrr}
      a_{0} + a_y + a_z + a_{yy} + a_{zz} + a_{yz}\\
      b_{0} + b_y + b_z + b_{yy} + b_{zz} + b_{yz}\\
      c_{0} + c_y + c_z + c_{yy} + c_{zz} + c_{yz}
   \end{array} \right)
\end{align}
\end{subequations}
Then we apply the following finite difference
\begin{align*}
  2 \der{\Phi}{x} = - \Phi(0,0,0) + \Phi(1,1,0) + \Phi(1,0,1) - \Phi(0,1,1) \\
  2 \der{\Phi}{y} = - \Phi(0,0,0) + \Phi(1,1,0) - \Phi(1,0,1) + \Phi(0,1,1) \\
  2 \der{\Phi}{z} = - \Phi(0,0,0) - \Phi(1,1,0) + \Phi(1,0,1) + \Phi(0,1,1)
\end{align*}
to each of strain rate componenets $S_{\alpha\beta}$ and obtain
\begin{subequations}\label{eq:compact_3D_s}
\begin{alignat}{2}
  2 \der{\sxx}{x} &= 4a_{xx} &&= - \sxx(H) + \sxx(K) + \sxx(M) - \sxx(N) \\
  2 \der{\sxx}{y} &= 2a_{xy} &&= - \sxx(H) + \sxx(K) - \sxx(M) + \sxx(N) \\
  2 \der{\sxx}{z} &= 2a_{zx} &&= - \sxx(H) - \sxx(K) + \sxx(M) + \sxx(N) \\
  2 \der{\syy}{x} &= 2b_{xy} &&= - \syy(H) + \syy(K) + \syy(M) - \syy(N) \\
  2 \der{\syy}{y} &= 4b_{yy} &&= - \syy(H) + \syy(K) - \syy(M) + \syy(N) \\
  2 \der{\syy}{z} &= 2b_{yz} &&= - \syy(H) - \syy(K) + \syy(M) + \syy(N) \\
  2 \der{\szz}{x} &= 2c_{zx} &&= - \szz(H) + \szz(K) + \szz(M) - \szz(N) \\
  2 \der{\szz}{y} &= 2c_{yz} &&= - \szz(H) + \szz(K) - \szz(M) + \szz(N) \\
  2 \der{\szz}{z} &= 4c_{zz} &&= - \szz(H) - \szz(K) + \szz(M) + \szz(N) \\
  2 \der{\sxy}{x} &= a_{xy} +2b_{xx} &&= - \sxy(H) + \sxy(K) + \sxy(M) - \sxy(N) \\
  2 \der{\sxy}{y} &=2a_{yy} + b_{xy} &&= - \sxy(H) + \sxy(K) - \sxy(M) + \sxy(N) \\
  2 \der{\sxy}{z} &= a_{yz} + b_{zx} &&= - \sxy(H) - \sxy(K) + \sxy(M) + \sxy(N) \\
  2 \der{\syz}{x} &= b_{zx} + c_{xy} &&= - \syz(H) + \syz(K) + \syz(M) - \syz(N) \\
  2 \der{\syz}{y} &= b_{yz} +2c_{yy} &&= - \syz(H) + \syz(K) - \syz(M) + \syz(N) \\
  2 \der{\syz}{z} &=2b_{zz} + c_{yz} &&= - \syz(H) - \syz(K) + \syz(M) + \syz(N) \\
  2 \der{\sxz}{x} &= a_{zx} +2c_{xx} &&= - \sxz(H) + \sxz(K) + \sxz(M) - \sxz(N) \\
  2 \der{\sxz}{y} &= a_{yz} + c_{xy} &&= - \sxz(H) + \sxz(K) - \sxz(M) + \sxz(N) \\
  2 \der{\sxz}{z} &=2a_{zz} + c_{zx} &&= - \sxz(H) - \sxz(K) + \sxz(M) + \sxz(N)
\end{alignat}
\end{subequations}
By this we get the other 18 equations, thus complete the linear equation system
for the 30 unknown coefficients in the quadratic polynomial.
The coefficients of the quadratic terms ($a_{xx}$, $a_{xy}$, .etc) can be first
solved by Eq.~\eqref{eq:compact_3D_s}.
Then the remaining coefficients can solved by Eq.~\eqref{eq:compact_3D_u}.
Including the above algorithm, the interpolation procedure include the following
steps:
\begin{enumerate}[Step 1.]
  \item For each ghost element, calculate the velocity, pressure, $f^{neq}$
        and strain rate from the source elements.
  \item Interpolate velocity quadraticlly, pressure and $f^{neq}$ linearly.
  \item Calculate $f^{eq}$ from velocity and pressure.
  \item Rescale $f^{neq}$ by
          $\omega_{c} f_c^{neq} = \omega_{f} f_f^{neq}$
  \item Assign the \pdf of ghost element by $f = f^{eq} + f^{neq}$.
\end{enumerate}
\gff{} element always take the fluid elements as its sources, whereas
\gfc{} usually itself requires data from ghostFromFiner elements as
part of its sources.
Hence, interpolation of ghostFromFiner is always performed before
ghostFromCoarser to provide valid values for the latter one.

Temporal interpolation is not required in the current study as two layers of
ghost elements are constructed around fluid elements on finer levels.
During the synchronous time step, both layers get filled up by interpolation.
The outer layer becomes invalid after the next asynchronous time step, but the
inner ring still provides valid values for the streaming step.
This idea can also be used for diffusive scaling
(i.e. $\Delta t \propto \Delta x^2$), where four time steps need
to be performed on the finer level for each step on the coarser level.
Thus, four layers of ghost elements have to be constructed in this case.

\subsection{Boundary conditions}

For the flow over cylinder test case, the inflow, outflow ans no-slip wall
boundary conditions are needed.
The velocity bounce-back boundary condition~\cite{ladd1994numericalp1}
was applied at inlet,
\begin{equation}
  f_{\bar i}(\mathbf{x}_f,t+\dt) = f_{i}(\mathbf{x}_f,t) - 2f_{i}^{eq-}(\mathbf{x}_b,t)
\end{equation}
where $f_{i}^{eq-}$ is the antisymmetric equilibirium function and computed as
\begin{equation}
  f_{i}^{eq-} = 3 \omega \density(\mathbf{c}_i \cdot \vecvel).
\end{equation}
For outflow, an extrapolation algorithm combined with a pressure correction on
normal direction~\cite{junk2011asymptotic} was applied,
\begin{equation}
  f_{\bar i}(\mathbf{x}_f,t+\dt) = \left\{ \begin{array}{ll}
    f_i^{eq}(\density,\vecvel) + f_{\bar i}^{neq}, & \mathbf{c}_i = - \mathbf{n} \\
    2f_i(\mathbf{x}_f-\mathbf{n}) - f_i(\mathbf{x}_f-2\mathbf{n}), & \mathbf{c}_i \neq - \mathbf{n}
    \end{array} \right.
\end{equation}
where $\mathbf{n}$ is the outer normal direction of the boundary.
The no-slip wall boundary condition is based on the linear
interpolated bounce-back scheme proposed by Bouzidi~\cite{bouzidi2001:momentum-t},
where the distance between the actuall wall position and the
adjecent fluid element is taken into account to obtain more accurate results.
To evaluate the forces acting on the object, the momentum exchange
method~\cite{bouzidi2001:momentum-t} was implemented.
In this method, the change of distribution function is integrated over the
surface of a solid body
\begin{equation}
  \mathbf{F} = \sum_{\mathbf{x}_b} \sum_{i \in V} \mathbf{c}_i 
  [ f_{i}^{*}(\mathbf{x}_b,t)+f_{\bar i}(\mathbf{x}_b,t+\Delta t) ]
\end{equation}
where the position vector $\mathbf{x}_b$ denotes those fluid elements
intersecting the solid body and $V$ is the set of intersecting discrete
velocities.

\section{Octree based implementation}
\label{sec:implementation}

In this section, we describe the implementation of the Lattice Botlzmann models
aiming for the large parallel systems.
The growing number of computing units as well as the increasingly complex
character of simulations makes an efficient implementation a diffucult task.
Besides the standard computation and communication, the imbalance introduced by
mesh refinement and the interpolation put another challenge on efficiency.
Strategies based on sparse matrix~\cite{schulz2002sparse},
hierarchical tree structure~\cite{Harald:2012aa,neumann2013:a-dynamic-,eitel2013lattice}
and block-patch decomposition~\cite{schonherr2011multi,godenschwager2013framework}
have been developed for either homogeneous or inhomogeneous grid
targeting on multi-core systems or GPGPU platforms.
Next, we explain two of the major implementation features dealing with the
inhomogeneous grid challenges:
weight based domain decomposition and level-wise elements arrangment.

\subsection{Weight based domain decomposition}

The mesh representation in \musubi{} is based on the sparse octree data
structure.
The size of a mesh element is related to its level $L$ within the tree hierarchy by
\begin{equation}
  \Delta x = H / 2^L
\end{equation}
where $H$ is the maximum length of the enclosing computational domain.
Each element within the mesh is uniquely labeled by a single 8 byte interger
called \treeid{}.
The \treeid{} embeds the spatial and topological information about the element
within the global domain implicitly.
All the elements in the mesh can be uniquely linearized by traveling along a
space filling curve (e.g. z curve~\cite{morton1966computer} is used in \musubi{}).
As coarse and fine elements perform different number of iterations within the
same time unit, an integer workload weight $w_i$ is calculated for and assigned
to each element depending on its level
\begin{equation}
  w_i = 2^{L_i - L_{min}} \qquad i = 1,...,N
  \label{eq:weight}
\end{equation}
where $N$ is the number of elements and $L_{min}$ is the minimum level.
Then the cumulative weight $W_i$ is calculated along the element list
\begin{equation}
  W_i = \sum_{j=1}^{i}w_j \qquad i = 1,...,N
\end{equation}
by which the total weight of the whole domain and the average weight for
each partition is also obtained.
After that the linearized elements array is simply divided according the
cumulative weight so that each process get the same amount of cumulative weight
(i.e. workload).
This procudure allows an automatic domain partition for arbitrarily complex
and locally refined meshes.

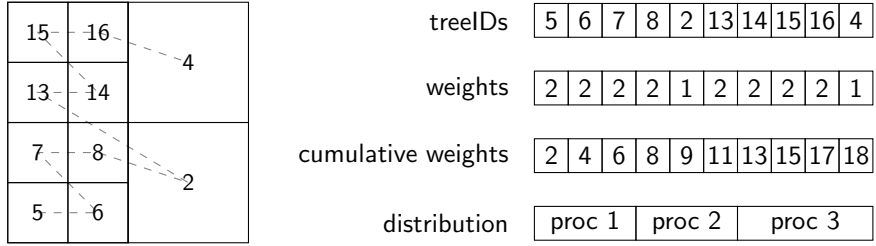
\begin{figure}
\centering
  \begin{subfigure}{.3\textwidth}
    \begin{tikzpicture}[scale=1.6,font=\sffamily]
    \tikzstyle{every label}=[font=\bfseries]
    \foreach \x in {0,1}
        \foreach \y in {0,1}
            \draw[black] (\x,\y) rectangle ++(1,1);
    \foreach \x in {0,1}
        \foreach \y in {0,1,2,3}
            \draw[black] (\x*0.5,\y*0.5) rectangle ++(0.5,0.5);
    \node [scale=0.9] at (0.25,0.25)   {5};
    \node [scale=0.9] at (0.75,0.25)   {6};
    \node [scale=0.9] at (0.25,0.75)   {7};
    \node [scale=0.9] at (0.75,0.75)   {8};
    \node [scale=0.9] at (0.25,0.25+1) {13};
    \node [scale=0.9] at (0.75,0.25+1) {14};
    \node [scale=0.9] at (0.25,0.75+1) {15};
    \node [scale=0.9] at (0.75,0.75+1) {16};
    \node [scale=0.9] at (1.5,0.5)     {2};
    \node [scale=0.9] at (1.5,1.5)     {4};
    \draw [dashed,opacity=0.5]        (0.25,0.25)   --
                      (0.75,0.25)   --
                      (0.25,0.75)   --
                      (0.75,0.75)   --
                      (1.5,0.5)     --
                      (0.25,0.25+1) --
                      (0.75,0.25+1) --
                      (0.25,0.75+1) --
                      (0.75,0.75+1) --
                      (1.5,1.5);
    \end{tikzpicture}
    \caption{Quadtree mesh}
    \label{fig:quadtree}
  \end{subfigure}
  \begin{subfigure}{.6\textwidth}
    \begin{tikzpicture}[scale=0.45,font=\sffamily]
      \foreach \x in {0,...,9}
          \draw[black] (\x-0.5,-0.5) rectangle ++(1.0,1.0);
      \node[left] at (-1,0) {treeIDs};
      \node[black] at (0,0) {5};
      \node[black] at (1,0) {6};
      \node[black] at (2,0) {7};
      \node[black] at (3,0) {8};
      \node[black] at (4,0) {2};
      \node[black] at (5,0) {13};
      \node[black] at (6,0) {14};
      \node[black] at (7,0) {15};
      \node[black] at (8,0) {16};
      \node[black] at (9,0) {4};
      \foreach \x in {0,...,9}
          \draw[black] (\x-0.5,-2.5) rectangle ++(1.0,1.0);
      \node[left] at (-1,-2.0) {weights};
      \node[black] at (0,-2.0) {2};
      \node[black] at (1,-2.0) {2};
      \node[black] at (2,-2.0) {2};
      \node[black] at (3,-2.0) {2};
      \node[black] at (4,-2.0) {1};
      \node[black] at (5,-2.0) {2};
      \node[black] at (6,-2.0) {2};
      \node[black] at (7,-2.0) {2};
      \node[black] at (8,-2.0) {2};
      \node[black] at (9,-2.0) {1};
      \foreach \x in {0,...,9}
          \draw[black] (\x-0.5,-4.5) rectangle ++(1.0,1.0);
      \node[left] at (-1,-4.0) {cumulative weights};
      \node[black] at (0,-4.0) {2};
      \node[black] at (1,-4.0) {4};
      \node[black] at (2,-4.0) {6};
      \node[black] at (3,-4.0) {8};
      \node[black] at (4,-4.0) {9};
      \node[black] at (5,-4.0) {11};
      \node[black] at (6,-4.0) {13};
      \node[black] at (7,-4.0) {15};
      \node[black] at (8,-4.0) {17};
      \node[black] at (9,-4.0) {18};
      \node[left] at (-1,-6.0) {distribution};
      \draw[black] (-0.5,-6.5) rectangle ++(3.0,1.0);
      \draw[black] ( 2.5,-6.5) rectangle ++(3.0,1.0);
      \draw[black] ( 5.5,-6.5) rectangle ++(4.0,1.0);
      \node[black] at (1,-6.0) {proc 1};
      \node[black] at (4,-6.0) {proc 2};
      \node[black] at (7.5,-6.0) {proc 3};
    \end{tikzpicture}
    \caption{Weights calculation and elements distribution}
    \label{fig:level}
  \end{subfigure}
  \caption{Domain decomposition of a simple two dimensional quadtree mesh.
(a) Each element is labeled by its \treeid{}. Dashed line indicates the z curve.
(b) Elements are sorted along the z curve and distributed among processes such
that total workload is averaged.}
  \label{fig:partition}
\end{figure}

An illustration of the above procudure by means of
a simple mesh example in two dimensions is shown in Fig.~\ref{fig:partition}.
The whole computational domain is considered as the root element with \treeid{}
0 on level 0.
The left half of the domain is refined unitl level 2, while the right half is
refined at level 1 as shown in Fig.~\ref{fig:quadtree}.
The number within each element is its \treeid{}.
All the elements are sorted according the z curve~\cite{morton1966computer}
(indicated by the dash line),
which results in a linearized array containing the \emph{treeIDs}
as shown in Fig.~\ref{fig:level}.
Then the weight and the cumulative weight is calculated for each element.
Assuming there are 3 processes and the total weight in this exmaple is 18, thus
the average weight for each partition is 6.
Finally the element list is divided such that each process has roughly the same
workload.

\subsection{Level-wise elements arrangment}

Computation kernel is the most time comsuming routine followed by interpolation
and communication.
Thus the purpose of elements arrangment is to facilitate these hot spots.
After loading mesh file, \musubi{} create level-wise lists of elements to allow
uniform operations on elements within each list (i.e. elements with the same
size).
In each list, fluid elements are first stored, then followed by ghost elements
and last by halo elements (i.e. elements from other processes).
Additionally, elements belonging to each type are sorted following the space
filling curve order to maxmize cache utilization.
Within the main iteration loop, all three types of elements are fed into
computation kernel in which advection and streaming steps are merged into one.
Then ghost elements as well as their depending source elements are fed into
the interpolation routine.
At last, halo elements are fed into communication routine where their values get
filled up from remote processes.
Such level-wise arrangment and element type classification allows for a
simple element loop treatment within the solver while hiding the rather
complex mesh behind.
Moreover, the chance of being able to perform optimization and
vectorization by compiler is also increased.

\section{Results}
\label{sec:results}

\subsection{Taylor-Green vortex}
First, we examined the convergence behaviour of the interpolation scheme based on
compact stencil using the 2D Taylor-Green vortex flow.
This unsteady and spatially periodic flow problem has an analytical solution for
the incompressible NS equations, thus serves as a popular test case~\cite{brachet1983small,kruger2010second,geier2015cumulant,fakhari2015numerics}.

The computational domain is a periodic square of size $0 \le \vecx \le 2\pi$
without any boundary condition.
The Reynolds number is defined by $Re = 1 / \nu$ and
was set to be $25$ in the present study.
The velocity $\vecvel$, pressure $\pressure$ and strain rate $S$ fields are given as
\begin{subequations}
  \begin{align}
  \vecvel(\vecx,t) &=
  u_0 \left( \begin{array}{rr}
      - \cos(x) \sin(y)\\
        \sin(x) \cos(y)
      \end{array} \right) e^{-t/t_D} \\
  p(\vecx,t) &= p_0 - \density\frac{u_{0}^{2}}{4}
  \left( \cos(2x) + \cos(2y) \right) e^{-2t/t_D} \\
  S_{xx}(\vecx,t) &= -S_{yy}(\vecx,t) = u_0 \sin(x) \sin(y) e^{-t/t_D} \\
  S_{xy} &= 0
  \end{align}
  \label{eq:tgv}
\end{subequations}
$t_D = 1 / 2\nu$ is the vortex decay time.
$u_0 = 1$ and $p_0 = 0$ are reference constants without hydrodynamic significance.
The $f_i$ are initialised by summing the equilibirium part $f^{eq}$,
cf. Eq.~\eqref{eq:feq}, and the non-equilibirium part
\begin{equation}
  f^{neq}_i \approx -\frac{w_i}{2c_s^2 \rho \nu \omega} \mathbf{Q}_i : S.
\end{equation}
The required macroscopic quantities are taken from Eq.~\eqref{eq:tgv}.
The numerical results are evaluated at $t=t_D$.
The error between simulation results and analytical solutions are evaluated
over the global domain in the form of l2-norm
\begin{equation}
  \epsilon_\phi(t) =
  \sqrt{\frac{\sum_x[\phi_s(\vecx,t)-\phi_a(\vecx,t)]^2}
             {\sum_x[\phi_a(\vecx,t)]^2}
       }
\end{equation}
where $\phi$ can be $u_x$ and $S_{xx}$ and indexes $s$ and $a$ denote
simulation and analytical results, respectively.
Simulations were performed with consecutively refined mesh ($L = 4,5,6,7$).
For each case, the left half of the domain was refined at level $L$, while the
right half was refined at one level higher.

\begin{figure}[h]
  \centering
  \includegraphics[width=0.7\textwidth]{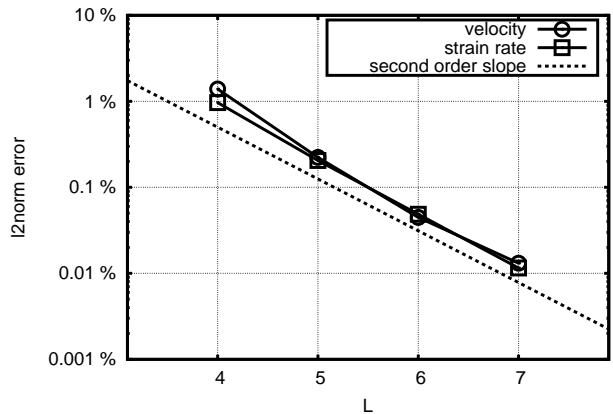}
  \caption{Convergence of velocity and strain rate for the Taylor-Green vortex
    test case. In each case, the left half domain was refined at level $L$,
    while the right half was refined at one level higher.}
  \label{fig:tgv_conv}
\end{figure}

As can be seen in Fig.~\ref{fig:tgv_conv}, the velocity shows a second order
convergence behaviour as expected.
Moreover, strain rate also presents almost the same convergence even though only
linear interpolation was used for $f_{neq}$ from which strain rate is computed.
This might be due to the fact that strain rate information was partially used
during the velocity interpolation process.

\subsection{Flow over cylinder at $Re=20,100$}

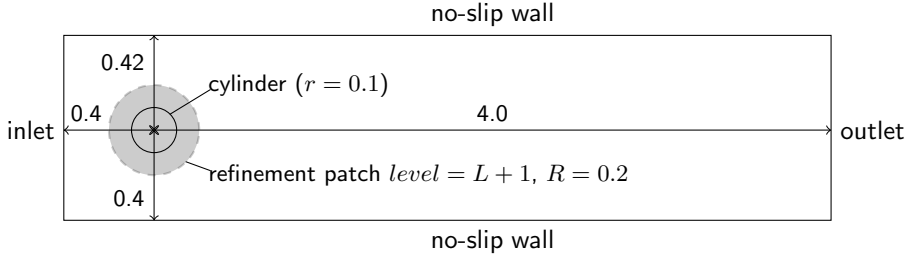
\begin{figure}
\centering
\begin{tikzpicture}[scale=0.3,font=\sffamily]
  \tikzstyle{every label}=[font=\bfseries]
  \draw [<->] (-4,0) -- (0,0);
  \draw [<->] (0,0) -- (30,0);
  \draw [<->] (0,-4.0) -- (0,0);
  \draw [<->] (0,0) -- (0, 4.2);
  \node [above]      at (15, 0) {\small 4.0};
  \node [left,above] at (-3, 0) {\small 0.4};
  \node [left]       at ( 0, 3) {\small 0.42};
  \node [left]       at ( 0,-3) {\small 0.4};
  \draw (-4,-4) rectangle (30,4.2);
  \node [left]  at (-4,0) {inlet};
  \node [right] at (30,0) {outlet};
  \node [above] at (15, 4.2) {no-slip wall};
  \node [below] at (15,-4.0) {no-slip wall};
  \draw ( 0, 0) circle [radius=1];
  \draw [-] (0.7,0.7) -- (2.5,1.8);
  \node [right] at (2, 2) {\small cylinder ($r=0.1$)};
  \draw [dashed,thick,fill=black,opacity=0.2] ( 0, 0) circle [radius=2];
  \draw [-] (1.4,-1.4) -- (2.5,-1.8);
  \node [right] at (2,-2) {\small refinement patch $level=L+1$, $R=0.2$};
\end{tikzpicture}
\caption{Geometry configuration of flow over cylinder for $Re=20,100$}
\label{fig:cyl_config}
\end{figure}

The simulation of flow over cylinder in 2D is a popular test case as it
investigate various flow features depending on the Reynolds number.
Detailed reference results can be found in~\cite{schafer1996benchmark,schonherr2011multi}
which involve results from not only LBM but also other numerical schemes.

The geometry configuration defined in~\cite{schafer1996benchmark} is shown in
Fig.~\ref{fig:cyl_config}.
The cylinder with radius of 0.1 is placed asymmetrically inside the channel
with the ratio betwee length and height equals to 5.
The outer ring region with a thickness of 0.1 around the cylinder was always refined
at one level higher (L+1) than the bulk area (L).
This setup was simulated with three meshes at different resolutions,
i.e. $D=40\Delta x_{L+1}, 80\Delta x_{L+1}, 160\Delta x_{L+1}$, repectively.
A parabolic velocity was applied at the inlet whereas outflow was set at the
outlet.
The Reynolds number is defined as $Re=D\bar u/\nu$, where $D$ is the
cylinder diameter, $\bar u$ is the mean velocity at inlet and $\nu$ is the
kinematic viscosity.
The drag and lift coefficients, $c_D$ and $c_L$, are calculated by
\begin{equation*}
  c_D = \frac{2F_x}{\density \bar u^2 A} \qquad \text{and} \qquad
  c_L = \frac{2F_y}{\density \bar u^2 A},
  \label{eqn:cdcl}
\end{equation*}
where $F_x$ and $F_y$ is the x- and y-component of the forces integrated over the
whole cylinder exerted by the fluid, $A = D \Delta x$ is the confronting area.
The Strouhal number is defined as $St = Df/u_{\infty}$, where $f$ is the
frequency of separation.
The flow at $Re=20,100$ was investigated seperately.
The $Re=20$ case is a steady flow, where drag and lift coefficients are
calcuated.
The other case is a unsteady flow resulting in a von K\'arm\'an vortex street,
where the maxminum value of drag and lift coefficient over one period
(i.e. $c_{Dmax}$ and $c_{Lmax}$)
as well as the $St$ number is calculated and compared with results in literature.

\begin{figure}
\centering
  \begin{subfigure}{.5\textwidth}
  \centering
  \includegraphics[width=.99\textwidth]{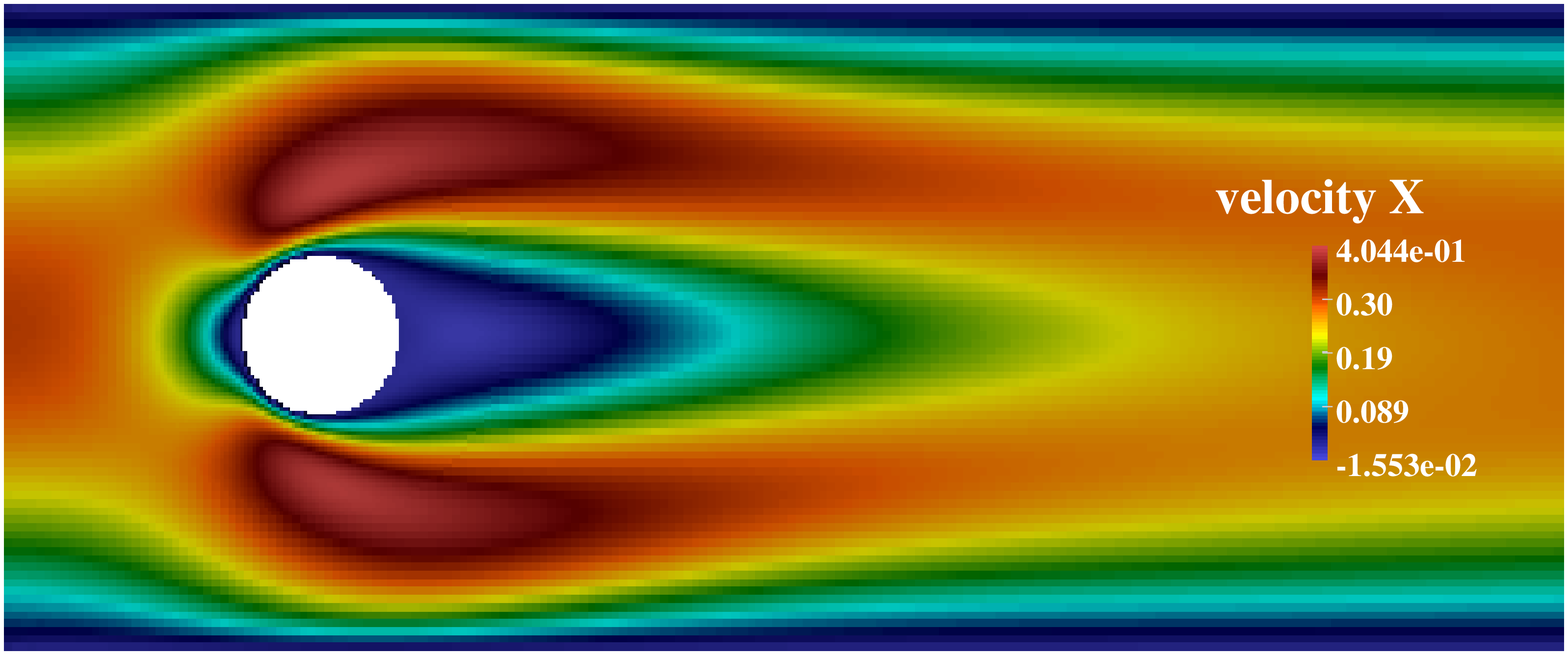}
  \caption{$Re=20$}
  \end{subfigure}%
  \begin{subfigure}{.5\textwidth}
  \centering
  \includegraphics[width=.99\textwidth]{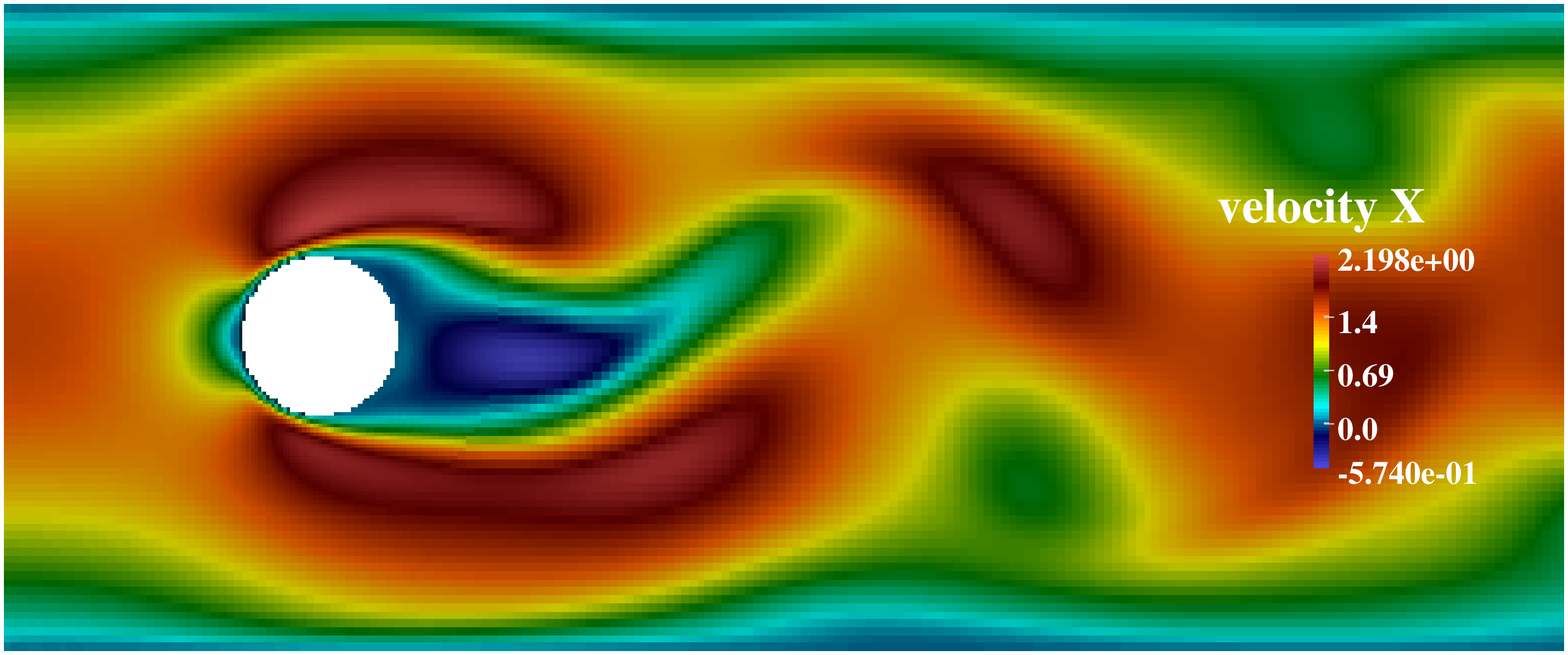}
  \caption{$Re=100$}
  \end{subfigure}
\caption{Velocity distribution for flow over cylinder at (a) $Re=20$ and (b)
  $Re=100$. The velocity magnitude is given by the color map.}
\label{fig:vel_cyl}
\end{figure}

The velocity distributions in Fig.~\ref{fig:vel_cyl} illustrate the transition
from the steady flow to the unsteady oscllating flow regime.
The simulaiton results are listed in Table \ref{table:cyl}. Good agreement
between simulated results and the ones from literature is obtained for both
flow problems.

\begin{table}[h!]
\caption{Simulation results of drag and lift coefficients, $c_D$ and $c_L$,
         for flow over cylinder at $Re=20,100$}
\centering
\begin{tabular}{lp{1.3cm}p{1.4cm}p{1.4cm}p{1.4cm}p{1.0cm}}
  \hline
            & \multicolumn{2}{l}{$Re=20$} & \multicolumn{3}{l}{$Re=100$} \\
              \cline{2-3} \cline{4-6}
  resolution & $c_D$ & $c_L$ & $c_{Dmax}$ & $c_{Lmax}$ & $St$ \\
  \hline
  $D= 40\Delta x_{L+1}$ & 5.745 & 0.01438 & 3.285 & 0.9751 & 0.3087\\
  $D= 80\Delta x_{L+1}$ & 5.664 & 0.01234 & 3.272 & 0.9934 & 0.3088\\
  $D=160\Delta x_{L+1}$ & 5.621 & 0.01146 & 3.245 & 0.9888 & 0.3034\\
  literature~\cite{schafer1996benchmark,crouse2003lattice,schonherr2011multi}
  & 5.49-5.627 & 0.0092-0.0119 & 3.22-3.2650 & 0.9492-1.0709 & 0.295-0.3076 \\
  \hline
\end{tabular}
\label{table:cyl}
\end{table}


\subsection{Flow over sphere at $Re=100$}

The flow over sphere benchmark problem was chosen to validate our approach for
the 3D case.
Similar to the flow over cylinder, the flow phenomena can be categorized into
different regimes based on the Reynolds number.
Previous studies~\cite{johnson1999flow} have shown that the flow is steady and
axisymmetric up to $Re \approx 210$.
In the present study, the flow at $Re=100$ is chosen to evaluate the refinement
mothod in three dimensions.
The computational domain has a size of
$[-15D,49D] \times [-15D,15D] \times  [-15D,15D]$.
Four levels of refinement are used, which are clusted around the sphere located
at $\mathbf{x} = (0,0,0)$.
Element size equals $\Delta x_{min} = D/32$ on the highest mesh level.
The drag coefficients $c_D$ are defined the same as Eq.~\eqref{eqn:cdcl},
where $A = \pi D^2/4$ for a sphere.
The simulated results of $c_D$ and $Lr/D$ (normarlized recirculation length)
are listed in Table~\ref{table:sph} and in good agreement with
those from the literature, proving the accurary of our approach in three
dimensions.

\begin{table}[h!]
\caption{Simulation results of drag coefficients $c_D$ and normalized
  recirculation length $Lr/D$ for flow over sphere at $Re=100$}
\centering
\begin{tabular}{lp{1.6cm}p{2cm}}
  \hline
          & $c_D$ & $Lr/D$ \\
  \hline
  present & 1.090 & 0.872 \\
  literature~\cite{johnson1999flow,hartmann2011strictly, eitel2011lattice}
  & 1.06-1.098 & 0.863-0.880 \\
  \hline
\end{tabular}
\label{table:sph}
\end{table}

\section{Conclusion}
\label{sec:conclusion}

In this paper, we presented a detailed LB implementation on non-uniform sparse octree
data structure.
Our method makes use of the strain rate information locally available to achieve
quadratic interpolation for the velocity.
This method requires only a minimum of four source elements from the adjecent
level for both 2D and 3D, thus is highly efficient especially for paralization.
To allow fluid elements to performance two consecutive advection steps,
two layers of ghost elements are introduced at the mesh level interface
and get filled up by interpolation during synchorous step.
Two of the main implementation features are explained in details: weight based
domain decomposition and level-wise elements arrangement.
These are specially designed to improve efficiency in parallel simulations.
Through the Taylor-Green vortex test case, the second order convergence was
achieved for both velocity and strain rate.
Moreover, laminar flow over cylinder in 2D and flow over sphere in 3D was
investigated, where drag and lift coefficients, Strouhal number and recirculation
length were calculated.
Good agreement between simulation results and those from literature was
observed, evidencing the accuracy of our method.

In the present study, region with mesh refinement was defined in the mesh
generation step and fixed during flow simulation.
The mesh adaptivity techniques are among our ongoing efforts.
A dynamic load balancing algorithm was also under investigation
to account for the altered mesh topology.

\section*{Acknowledgments}
We acknowledge PRACE for awarding us access to the Cray XC40 Hazel Hen system in
High Performance Computing Center Stuttgart (HLRS), Stuttgart, Germany, through
the project number 2730.

\bibliographystyle{elsarticle-num}
\bibliography{mybib}

\end{document}